%% file: main.tex
\documentclass[conference]{IEEEtran}

\usepackage{graphicx}
\usepackage{amsmath}
\usepackage{cite}
\usepackage{url}
\usepackage{tikz}
\usetikzlibrary{arrows.meta}
\usepackage[colorlinks=false,hidelinks]{hyperref}

\begin{document}

\title{Protocol Integration of Physical Layer Deception into EAP-TEAP Wi-Fi Authentication}

\author{
\IEEEauthorblockN{Moustafa Ibrahim$^{*}$, Bin Han$^{*}$, Hans D.\ Schotten$^{*\dagger}$}
\IEEEauthorblockA{$^{*}$RPTU Kaiserslautern-Landau, Kaiserslautern, Germany\\
$^{\dagger}$German Research Center for Artificial Intelligence (DFKI), Kaiserslautern, Germany}
}

\maketitle

\begin{abstract}
Credential-based Extensible Authentication Protocol (EAP) authentication
cannot distinguish a legitimate credential holder from an adversary
using compromised credentials. Physical Layer Deception (PLD)
complements credential-based authentication by exposing a deceptive
primary object over a primary transport while a separate recovery
object travels with differentiated reliability over a secondary
channel. Existing PLD studies remain, to our knowledge, at the physical/link-model level; using PLD's activation/deactivation mechanism as an authentication gate creates an authentication-specific design requirement, since an all-inactive attempt would exercise no recovery path. We present a batched PLD-based re-verification step for
Enterprise Wi-Fi's TEAP/RADIUS/IEEE~802.11 authentication chain,
implemented end to end across the server, access point, and device in
the open-source hostap~2.12 codebase. Each attempt carries three
rounds, at least one active, with no dedicated activation flag. Across
four campaigns totaling 1593 attempts, the prototype evaluates batched
recovery behavior, rejects the implemented naive credential-bearing
attacker in all 30 attempts, measures successful-path latency, and
evaluates the security-reliability trade-off for one, two, and three
active rounds under two modeled recovery regimes. The evaluation
exercises the protocol and software-MAC behavior directly and analyzes
informed and retry-seeking attackers under the software recovery model.
\end{abstract}


\section{Introduction}
\label{sec:intro}

Enterprise wireless local area networks commonly employ IEEE~802.1X-based
port access control \cite{ieee8021x2020}, where a claimant's Extensible
Authentication Protocol (EAP) \cite{aboba2004eap} exchange is relayed
through an access point to a RADIUS \cite{rigney2000radius} server,
often via Tunnel EAP (TEAP) \cite{dekok2026teap}. Such schemes
authenticate whoever presents valid credentials: once
credentials are cloned, exfiltrated, or otherwise compromised, an
adversarial claimant becomes indistinguishable from the legitimate
device at the credential layer alone.

Physical-layer security complements credential-layer decisions with
channel/hardware evidence difficult for a remote party to reproduce
\cite{poor2017wireless}: channel-based authentication verifies
continued presence via time-variant channel characteristics
\cite{xiao2008physical}, and challenge-response variants such as
PHY-CRAM \cite{shan2013phycram} extend this to an explicit
exchange \cite{xie2021survey,wang2016physical}. Physical Layer
Deception (PLD) is a distinct, active mechanism: it exposes a deceptive
primary object over a primary transport while a separate,
differentiated-reliability recovery object lets a favorably
placed receiver recover the true value more reliably than a
disadvantaged one \cite{han2023nom,chen2025twc,han2025semantic,chen2025ofdm}.

To our knowledge, existing PLD studies do not address integrating
PLD's activation/deactivation mechanism into a running EAP stack.
Applying that mechanism as an authentication gate creates an
authentication-specific design requirement: at least one active round
must occur, since an all-inactive attempt would expose every true
challenge and never exercise the recovery path. The proposed scheme
satisfies this requirement with a batch of $L$ rounds and exactly $A$ active
positions, enforcing $A\ge1$. Integrating PLD into
EAP-TEAP is further a cross-layer placement problem: the primary
object must stay opaque to the access point inside the TEAP/TLS
tunnel, while the recovery object reaches the device over a separate
IEEE~802.11 path, correlated without a fixed arrival order, using only
existing TEAP, RADIUS, and IEEE~802.11 extensibility. The proposed scheme resolves this placement problem and addresses the design requirement described above. Our contributions are: (i) a
standards-aware cross-layer integration of PLD into
EAP-TEAP/RADIUS/IEEE~802.11, preserving TEAP opacity for primary
objects while exporting recovery objects over separate Category-127
MAC frames; (ii) an authentication-specific batched construction with
$L$ rounds and exactly $A$ active positions, $A\ge1$ in normal
operation, inactive litter, no activation flag, and all-round final
gating; (iii) an end-to-end hostap~2.12 implementation across the
server, access point, and device, evaluated under
\texttt{mac80211\_hwsim} through four campaigns totaling 1593 attempts
covering batched recovery behavior, a naive-attacker baseline,
latency, and the security-reliability trade-off; and (iv) an
analytical informed-attacker and reliability treatment, including an
adaptive retry-seeking attacker that exploits repeated observations of
the same recovery object, deriving false-accept, PLD-induced
false-reject, and retry/security trade-off behavior, with overhead
derived from the implemented message formats and a proposed bounded
fresh-session retry policy.

The remainder of the paper is organized as follows.
Section~\ref{sec:design} presents the proposed design; Section~\ref{sec:attacker}
analyzes attacker and reliability behavior; Section~\ref{sec:eval}
describes the implementation and evaluation; and
Section~\ref{sec:discussion} concludes.

\section{PLD Authentication Model and Batched Re-Verification Design}
\label{sec:design}

\subsection{PLD Mechanism}
\label{sec:pldmech}

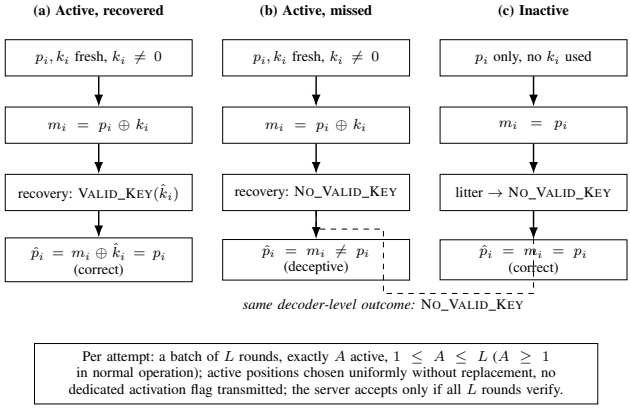
\begin{figure}[t]
\centering
\resizebox{0.94\columnwidth}{!}{\input{fig_pld_mechanism.tex}}
\caption{PLD re-verification decision semantics for one round under
the three decoder cases.}
\label{fig:pldmech}
\end{figure}

PLD pairs a primary object, carried over a primary transport, with a
recovery object carried over a separate transport with differentiated
reliability. The proposed re-verification applies this to a per-round true challenge $p_i$: on an
\emph{active} round the server exposes $m_i=p_i\oplus k_i$ for a fresh
nonzero key $k_i$ (so $m_i\ne p_i$), paired with a genuine key-bearing
recovery object; on an \emph{inactive} round it exposes $m_i=p_i$
directly, paired with a same-length but non-key-bearing \emph{litter}
object, so the object's presence alone never reveals inactivity. A
receiver that recovers a valid key (\textsc{Valid\_Key}) reconstructs
$\hat{p}_i=m_i\oplus\hat{k}_i=p_i$; otherwise it falls back to
$\hat{p}_i=m_i$ (\textsc{No\_Valid\_Key}), which is correct on an
inactive round but \emph{wrong} on an active one whose recovery failed.
Fig.~\ref{fig:pldmech} illustrates all three outcomes.

An inactive litter object that fails validation and an unrecovered
active recovery object are both mapped to \textsc{No\_Valid\_Key},
invoking the same fallback $\hat{p}_i=m_i$. The random-guess baseline of Section~\ref{sec:fa} assumes that no side
information distinguishes the two cases. An intact active recovery
object instead lets a party confirm activation through the public
validity check of Section~\ref{sec:construction}.

An inactive round exposes $m_i=p_i$, so an all-inactive attempt would
bypass the recovery path. We therefore select exactly $A\geq1$ active
positions uniformly without replacement and transmit no activation
flag; all $L$ responses must verify. Randomized placement contributes
the combinatorial uncertainty in Eq.~\eqref{eq:rg-accept} when
unresolved active positions must be inferred, while the Perception result $r_E^A$ and Bob's marginal
acceptance under i.i.d.\ recovery do not depend on placement.

\subsection{Threat Model}
\label{sec:threat}

The proposed scheme targets a claimant that already holds compromised client
credential material and can complete ordinary EAP-TEAP/EAP-TLS
authentication unassisted; it adds a PLD-oriented post-credential
re-verification step gating final EAP-Success. TLS protects $m_i$
from third-party observers, but the credential-bearing claimant
terminates its own TEAP/TLS session and therefore observes $m_i$; the
proposed scheme discriminates claimants through recovery-path reliability. We denote the legitimate receiver by Bob and the 
credential-bearing claimant by Eve, with per-round 
recovery probabilities $r_B$ and $r_E$, respectively. We assume a Kerckhoffs-style adversary that knows
the full construction, including $L$ and the active-set distribution,
but not its per-attempt realization. The server and access point are
trusted; the backend RADIUS path is outside this adversary's
observation, integrity-protected by the RADIUS Message-Authenticator.
For an inactive round $m_i$ already equals the true value $p_i$; for an active round,
an adversary that reliably obtains both $m_i$ and that round's
recovery object recovers $p_i$. Section~\ref{sec:attacker} quantifies what an
adversary without such reliable recovery can still achieve.

\subsection{Batched Construction and Protocol Mapping}
\label{sec:construction}

Mapping PLD's primary/recovery split onto batched EAP-TEAP requires
keeping TEAP/TLS opacity for primary objects, exporting each round's
recovery object to the access point as its own access-point-to-device MAC frame correlated
per round by TxnID/Round with no fixed arrival order, guaranteeing
$A\ge1$, and gating authentication on all $L$ rounds verifying, using
only existing TEAP/RADIUS/IEEE~802.11 extensibility.

Each round $i$ is either active or inactive, chosen by the
server:
\begin{equation}
  m_i = \begin{cases} p_i\oplus k_i, & i\ \text{active},\\ p_i, & i\ \text{inactive}, \end{cases}
  \label{eq:mi}
\end{equation}
where $p_i$ is drawn uniformly from the 128-bit strings and, for an
active round, $k_i$ uniformly from the nonzero 128-bit strings, so
$m_i\ne p_i$. Conditioned on an active $m_i$, $p_i$ remains one of
$2^{128}-1$ other values; Section~\ref{sec:fa} uses this quantity for
the blind-guess baseline.

Every round carries a 32-byte recovery-path object; an active
round's is the codeword
\begin{equation}
\begin{aligned}
  q_i &= k_i \,\|\, t_i,\\
  t_i &= \mathrm{Trunc}_{128}\big(\mathrm{SHA256}(
    \ell_Q \,\|\, \mathrm{TxnID}_{\mathrm{BE64}}\\
  &\qquad \,\|\, \mathrm{Round}_{i,\mathrm{BE16}} \,\|\, k_i)\big),
\end{aligned}
\label{eq:qcode}
\end{equation}
where $\ell_Q$ is a fixed public recovery-code domain-separation
label, $\mathrm{BE64}$ and $\mathrm{BE16}$ denote 64- and 16-bit
big-endian encodings, respectively, and $k_i$ is included both as-is
and inside $t_i$'s hash input. Decoding Eq.~\eqref{eq:qcode} takes the first 128 bits as a candidate key $\hat{k}_i$ and
recomputes $t_i$: a match yields \textsc{Valid\_Key}$(\hat{k}_i)$, a
mismatch \textsc{No\_Valid\_Key}.

This check is public and unkeyed, so
any party holding an intact active $q_i$ and its TxnID/Round can
recompute it. The public check establishes codeword validity only.
An inactive round's object is \emph{litter}: an
independently random 32-byte candidate checked against the same test;
one redraw is attempted on accidental validation, and a second
accidental validation fails batch construction safely rather than
deliver litter that decodes \textsc{Valid\_Key}.

No dedicated active/inactive flag or activation-mask field is
transmitted; the active-set realization is maintained as server state; the same public-validity-check ambiguity of Section~\ref{sec:pldmech}
applies to this wire encoding of Eq.~\eqref{eq:mi}.

A legitimate receiver reconstructs
\begin{equation}
  \hat{p}_i =
  \begin{cases}
    m_i \oplus \hat{k}_i, & \textsc{Valid\_Key},\\
    m_i, & \textsc{No\_Valid\_Key},
  \end{cases}
  \label{eq:phat}
\end{equation}
which equals $p_i$ on an active round decoded \textsc{Valid\_Key} with
$\hat{k}_i=k_i$, and trivially on any inactive round.

Let $L$ be the batch size; the server selects exactly $A$ of the $L$
rounds as the active set $\mathcal{A}$, by the placement rule of
Section~\ref{sec:pldmech}. Normal operation enforces $1\le A\le L$. At
$A=0$, every $m_i=p_i$, so the batch exposes every true value directly
and a claimant needs no recovery; $A=0$ is retained only as the
explicitly authorized all-inactive diagnostic condition of
Section~\ref{sec:campaign}, not during normal operation. The device
returns
\begin{equation}
\begin{aligned}
  R_i = \mathrm{HMAC\text{-}SHA256}\big(\hat{p}_i,\ &\ell_R\,\|\,\mathrm{TxnID}_{\mathrm{BE64}}\\
  &\,\|\,\mathrm{Round}_{i,\mathrm{BE16}}\,\|\,L_{\mathrm{BE16}}\big),
\end{aligned}
\label{eq:hmac}
\end{equation}
where $\ell_R$ is a fixed public response domain-separation label and
$L_{\mathrm{BE16}}$ encodes the batch size $L$; the server
independently recomputes this per round, keyed on its own $p_i$, accepting the batch only if all $L$ rounds occur exactly once and every
$R_i$ of Eq.~\eqref{eq:hmac} matches.

\begin{figure}[t]
\centering
\resizebox{0.94\columnwidth}{!}{\input{fig_w1v2_flow.tex}}

\caption{Batched PLD re-verification message flow for a batch of $L$
rounds ($L\le3$ evaluated). The server retains the active-set
realization, the access point relays the TEAP-protected primary
objects, and each recovery object is forwarded in a per-round
Category-127 Action frame.}
\label{fig:w1v2flow}
\end{figure}
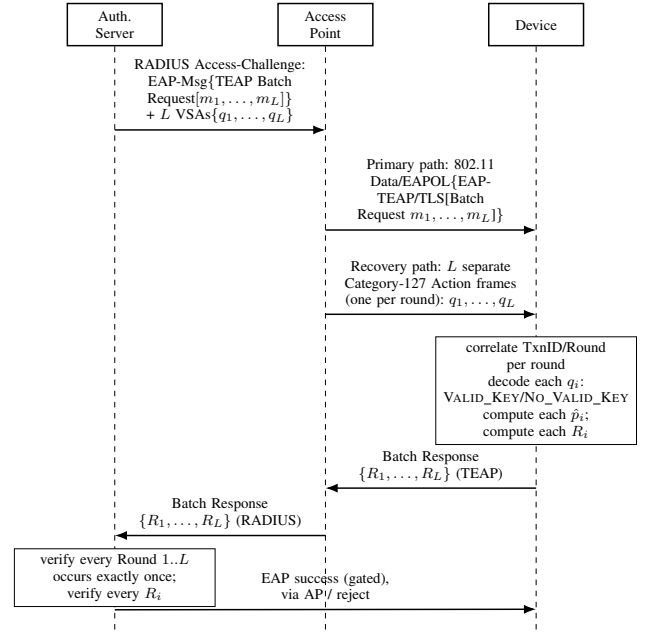

One TEAP Vendor-Specific TLV carries all $L$ values $m_1,\dots,m_L$
from server to device, opaque to the access point: Type~7, under
RFC~9930 \cite{dekok2026teap}, where we use Vendor-Id
\texttt{0x00FFFFFE} as a laboratory placeholder. Each $q_i$ instead
travels as
its own RADIUS Type-26 VSA \cite{rigney2000radius,aboba2003radiuseap},
all $L$ attached to the same Access-Challenge. After the access point
verifies the RADIUS Message-Authenticator, it extracts the $L$
recovery VSAs and forwards each $q_i$ as its own Category-127
IEEE~802.11 Vendor-Specific Action frame \cite{ieee80211_2024}. Category-127
provides the existing vendor-specific carrier; the semantics of the
proposed scheme and the laboratory Vendor-Id remain
implementation-specific. A batch thus
produces one Batch
Request, $L$ VSAs, $L$ Action frames, and one Batch Response, correlated by TxnID/Round, reflecting
TEAP-opacity preservation while exposing recovery objects separately, as illustrated in Fig.~\ref{fig:w1v2flow}. 

The proposed re-verification is
an auxiliary post-inner-method TEAP Phase-2 TLV exchange that preserves the
standard TEAP key-derivation procedure and operates alongside
Crypto-Binding before the final Result TLV, following RFC~9930
Appendix~C.8 \cite{dekok2026teap}.

After accepting the Batch Request, the device waits 2000~ms by default
for missing $q_i$; unresolved rounds then take the
\textsc{No\_Valid\_Key} branch of Eq.~\eqref{eq:phat}, correct only if
that round is inactive. EAP retransmission may resend the stored Batch
Request without regenerating its session-bound objects, and received
$q_i$ values remain cached by TxnID/Round. The access point does not re-send a recovery Action frame merely
because the stored EAP request is retransmitted; repeated
observations of the same object arise only from lower-layer
retransmissions.

Malformed, missing, duplicate, or
mismatched Batch Responses fail the batch. A new authentication
session generates a fresh TxnID, active set, $p_i$, $k_i$, $m_i$, and
$q_i$.

\section{Attacker and Reliability Analysis}
\label{sec:attacker}

\subsection{Receiver and Attacker Strategies}
\label{sec:strategies}

Given the construction above, the \emph{legitimate receiver} applies
Eq.~\eqref{eq:phat} exactly. The \emph{naive non-deceptive attacker},
evaluated in Section~\ref{sec:naive-eval}, sets $\hat{p}_i=m_i$ for
every round regardless of its own decoder's outcome, never using an
available key. The \emph{informed Perception attacker}, following the
semantic-PLD ``Perception'' strategy of \cite{han2025semantic}, applies
the receiver's rule with its own recovery probability $r_E$ in
place of $r_B$, using a correctly recovered key when available and
without requiring $\mathcal{A}$'s realization. The \emph{informed
random-guess attacker} also knows $L$, $A$, and the challenge-space
size but must simultaneously guess any unrecovered active round's
position and value. An attacker may also withhold its response or
exploit an available retry mechanism to obtain additional independent
observations of the \emph{same} recovery object before responding;
Section~\ref{sec:retry} analyzes this channel-independent
repeated-observation strategy. Channel-state-aware attempt selection
is outside the scope of the current analysis.

\subsection{False Acceptance and PLD-Induced False Rejection}
\label{sec:fa}

Let $d_p{=}128$ be the true-challenge bit-length. As
Section~\ref{sec:construction} defines, an active round's true $p_i$
is uniform, absent recovery, over the $2^{d_p}-1$ values other than
$m_i$, so a blind guess succeeds with probability $1/(2^{d_p}-1)$. For
this baseline, rounds without \textsc{Valid\_Key} are assumed
observationally indistinguishable: the attacker gets no delivery-state
signal separating received litter from a missed active object. For the
informed random-guess attacker, let $T$ be the number of active rounds
whose true value it obtains without guessing (e.g.\ with probability
$r_E$ each); conditioned on $T=t$ it must still guess the remaining
active positions and values:
\begin{equation}
  P_{\mathrm{RG}}(\text{accept}\mid T=t) =
  \frac{1}{\binom{L-t}{A-t}}
  \left(\frac{1}{2^{d_p}-1}\right)^{A-t}.
  \label{eq:rg-accept}
\end{equation}
Its unconditional false-accept probability, for i.i.d.\ per-round
recovery with probability $r_E$ ($T\sim\mathrm{Binomial}(A,r_E)$), is
\begin{equation}
  P_{\mathrm{FA}}^{\mathrm{RG}} =
  \sum_{t=0}^{A} \binom{A}{t} r_E^{t}(1-r_E)^{A-t}\,
  P_{\mathrm{RG}}(\text{accept}\mid T=t).
  \label{eq:fa-rg}
\end{equation}

Eq.~\eqref{eq:fa-rg} assumes the observational indistinguishability above;
side information identifying inactive positions instead conditions the
combinatorial factor on the remaining candidates, decreasing it toward
one.
Eq.~\eqref{eq:rg-accept} therefore defines the decoder-output
baseline under this ambiguity; side information narrows the remaining
candidate set.

The informed Perception attacker instead guesses nothing: writing $T$
for the number of active-round keys it correctly recovers, it falsely
accepts exactly when $T=A$,
\begin{equation}
  P_{\mathrm{FA}}^{\mathrm{Perception}} = \Pr(T=A),
  \label{eq:fa-perception-general}
\end{equation}
which, under i.i.d.\ recovery, reduces to
\begin{equation}
  P_{\mathrm{FA}}^{\mathrm{Perception}} = r_E^{A}.
  \label{eq:fa-perception}
\end{equation}
Symmetrically, with $r_B$ the legitimate receiver's per-round recovery
probability, the PLD-induced false-reject probability is
\begin{equation}
  P_{\mathrm{FR}}^{\mathrm{PLD}} = 1 - r_B^{A},
  \label{eq:fr-pld}
\end{equation}
conditioned on successful primary/TEAP delivery.
Eqs.~\eqref{eq:fa-perception}--\eqref{eq:fr-pld}
give the trade-off directly: $P_{\mathrm{FA}}^{\mathrm{Perception}}$
falls and $P_{\mathrm{FR}}^{\mathrm{PLD}}$ rises as $A$ increases,
raising the recovery burden on an informed attacker at the cost of
higher legitimate rejection unless $r_B$ is correspondingly high.

The $r_E^A$/$r_B^A$ closed forms hold only under i.i.d.\ per-round
recovery, specializing the general joint probability $\Pr(T=A)$ of
Eq.~\eqref{eq:fa-perception-general} for correlated or
position-dependent recovery; all false-accept expressions neglect the
negligible probability of an accidental HMAC-SHA256 match. 

\begin{figure*}[!t]
\centering
\includegraphics[width=\textwidth]{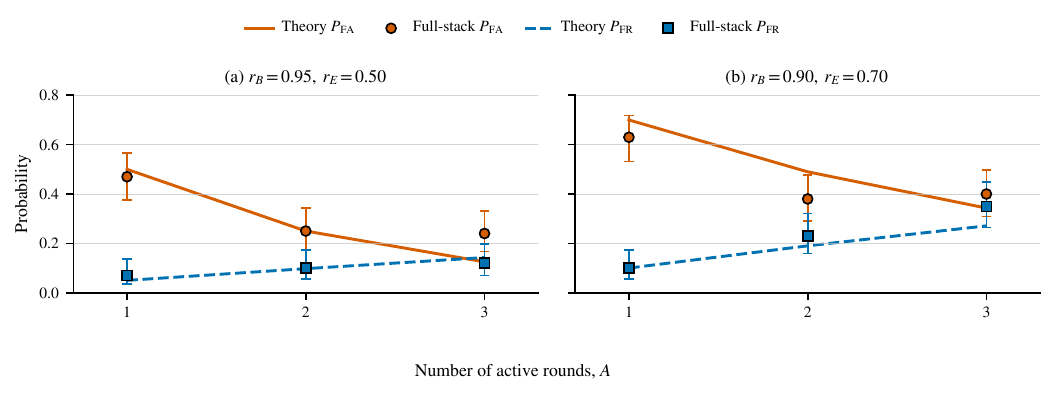}
\caption{Security-reliability trade-off versus the number of active
rounds $A$. Lines show the analytical $P_{\mathrm{FA}}=r_E^A$ and
$P_{\mathrm{FR}}=1-r_B^A$ characteristics; markers show full-stack
proportions from 100 attempts per condition with 95\% Wilson
confidence intervals.}
\label{fig:tradeoff}
\end{figure*}

\subsection{Repeated Observations and Retry Policy}
\label{sec:retry}

Because the PLD mechanism is public, Eve need not accept a single
observation: withholding a response or exploiting an available retry
mechanism can yield further independent observations of the
\emph{same} recovery object. The attempts considered here are
channel-independent: the retry decision does not condition on
instantaneous channel state.

Let $J$ denote the number of statistically independent effective
observations of the same active recovery object, including lower-layer
retransmission opportunities. The prototype enforces the batch and
failure-handling behavior of Section~\ref{sec:construction}, which
permits $J$ to exceed 1 through lower-layer retransmission. Under $J$
i.i.d.\ observations, the effective per-round recovery probability
becomes
\begin{equation}
  r_{E,J} = 1-(1-r_E)^J,
  \label{eq:rej}
\end{equation}
which tends toward 1 as $J$ grows even for small $r_E$. Under equal $J$
across the $A$ active rounds, the retry-aware false-accept probability
is
\begin{equation}
  P_{\mathrm{FA}}^{\mathrm{retry}} = \big[1-(1-r_E)^J\big]^A,
  \label{eq:fa-retry}
\end{equation}
the retry-aware analogue of Eq.~\eqref{eq:fa-perception}. The same
retry opportunity is available to the legitimate receiver: with
$r_{B,J}=1-(1-r_B)^J$,
\begin{equation}
  P_{\mathrm{FR}}^{\mathrm{retry}} = 1-r_{B,J}^{A}.
  \label{eq:fr-retry}
\end{equation}
Increasing $J$ raises both Bob's acceptance probability $r_{B,J}^A$ and
Eve's false-accept probability $P_{\mathrm{FA}}^{\mathrm{retry}}$ of Eq.~\eqref{eq:fa-retry}
together.

The failure handling of Section~\ref{sec:construction} limits same-object observations to lower-layer retransmissions: a failed batch terminates TEAP, whereas a new authentication session uses fresh
objects.

For $N$ independent fresh sessions with per-session false-accept
probability $P_{\mathrm{FA}}$,
\begin{equation}
  P_{\mathrm{FA},N} = 1-(1-P_{\mathrm{FA}})^N.
  \label{eq:fa-n}
\end{equation}
We therefore propose $N\leq N_{\max}$ followed by
backoff/rate-limiting and a suspicious-retry alarm. $N_{\max}$ remains
a deployment parameter because the same policy also limits legitimate
devices on persistently poor channels.

Randomized placement affects Eq.~\eqref{eq:rg-accept} only while
active positions remain unresolved. Eqs.~\eqref{eq:rej}--\eqref{eq:fa-n}
assume independent observations or sessions.

\section{Prototype and Evaluation}
\label{sec:eval}

\subsection{Implementation}
\label{sec:impl}

We implement the proposed scheme as additions to hostap~2.12: an Authentication Server
built on hostap's integrated RADIUS/EAP server, an Access Point
(\texttt{hostapd}), and a Device
(\texttt{wpa\_supplicant}), evaluated under the software MAC-level
simulator \texttt{mac80211\_hwsim}.
\label{sec:server}The server executes each attempt as one $L\le3$-round
batch, $L=3$ evaluated.

It assigns a fresh TxnID, selects
$A$ active positions by the placement rule of
Section~\ref{sec:pldmech}, builds each $m_i$/$q_i$, and sends one
Batch Request TLV and $L$ RADIUS VSAs on the same Access-Challenge. The
server precomputes every $R_i$ and withholds EAP-Success unless the
Batch Response contains every Round $1..L$ exactly once with every
$R_i$ matching; any malformed, missing, duplicate, or incorrect
material rejects the batch. We disable outer TEAP session resumption
so every measured authentication executes the full batch.

The access point performs ordinary relay, never terminating or
decrypting TEAP, so it never sees plaintext $m_i$. After the access
point verifies the RADIUS Message-Authenticator, it parses the $L$
VSAs and forwards each $q_i$, unexamined, as its own Category-127
Action frame. The access point forwards active codewords and litter
identically; validity checking occurs at the device.

\label{sec:device}The device correlates each round's $m_i$/$q_i$ by
TxnID/Round without assuming a fixed arrival order, including a $q_i$
that arrives before the Batch Request, which it caches pending the
request. It applies the public decoder to each available $q_i$, waits
up to the 2000-ms recovery deadline for any outstanding round,
computes all $L$ response tags, and returns them in one Batch
Response. Action-frame reception requires source-address/TxnID/Round
correlation and is registered through the nl80211 driver and exercised
through \texttt{mac80211\_hwsim}.

\subsection{Experimental Setup}
\label{sec:setup}

We run four campaigns under \texttt{mac80211\_hwsim}, all with $L=3$.
The 273-attempt recovery-model campaign covers the ten conditions in
Table~\ref{tab:w1v2campaign}. A 30-attempt naive-attacker campaign
uses an isolated test-only device binary with the server and access
point unchanged. A 90-attempt latency campaign uses 30 attempts each
of ordinary TEAP and the $A\in\{1,3\}$ full-recovery cases.
The 1200-attempt trade-off campaign covers two modeled recovery
regimes, $A\in\{1,2,3\}$, and Bob and Perception-Eve roles, with 100
attempts per condition. Each campaign uses a pre-generated fixed
schedule; degraded and trade-off recovery probabilities are realized
through controlled per-round recovery-object withholding. The shared
2000-ms deadline of Section~\ref{sec:construction} resolves rounds still
missing at expiry as \textsc{No\_Valid\_Key}. Reported latency covers
EAP-start-to-EAP-success on the accept path only; reject-path timing
is qualitative.

\subsection{Batched Recovery-Model Evaluation}
\label{sec:campaign}

\input{w1v2_campaign_table}

Table~\ref{tab:w1v2campaign} summarizes 273/273 complete attempts, with
no infrastructure failures or unexplained timeouts. Ordinary TEAP and
both full-recovery cases accept 30/30. Under degraded recovery, acceptance falls as the recovery probability decreases. The fixed and randomized $A=1$ conditions give the same
counts at both tested recovery probabilities, as expected under
i.i.d.\ recovery. The $A=0$ diagnostic accepts 3/3 with all
recovery objects withheld because every round is inactive, which
motivates $A\ge1$ in normal operation.

\subsection{Security-Reliability Trade-off Evaluation}
\label{sec:tradeoff}

To evaluate the trade-off in Eqs.~\eqref{eq:fa-perception} and
\eqref{eq:fr-pld}, we use two modeled recovery regimes,
$(r_B,r_E)=(0.95,0.50)$ and $(0.90,0.70)$, with $L=3$ and
$A\in\{1,2,3\}$. For each regime and $A$, 100 Bob attempts estimate
$P_{\mathrm{FR}}$ from the reject rate and 100 Perception-Eve attempts
estimate $P_{\mathrm{FA}}$ from the accept rate. Both roles use the
same device binary and decoding rule; only the modeled recovery probability differs.
Fig.~\ref{fig:tradeoff}
compares these proportions with the analytical curves and 95\% Wilson
confidence intervals. Increasing $A$ lowers analytical
$P_{\mathrm{FA}}$ and raises analytical $P_{\mathrm{FR}}$ in both
regimes. 

Finite recovery realizations from the pre-generated schedules
and randomized active-set placement produce sampling variation
around the nominal curves. All 1200 attempts reach an
authentication accept/reject decision, with zero infrastructure
failures or unexplained timeouts, and all 1200 outcomes match
the accept/reject decision predicted from the realized active set
and recovery-object delivery.

\subsection{Naive Credential-Bearing Attacker Baseline}
\label{sec:naive-eval}

We evaluate the naive attacker strategy of Section~\ref{sec:strategies}
over 30 attempts, using authorized laboratory credential material
at the $L=3,A=1$ randomized-position full-recovery configuration. All 3
recovery frames arrive in every attempt; the active round decodes
\textsc{Valid\_Key} and both inactive rounds decode
\textsc{No\_Valid\_Key}, confirming that a valid active-round key goes
deliberately unused. The HMAC-mismatch set matches exactly the active
round in every attempt, and all 30 attempts were rejected, with zero
unexpected accepts and no infrastructure failure. This result covers
this one strategy at this one configuration. The semantic-PLD
Perception strategy, which uses the available key, is analyzed in
Section~\ref{sec:fa}.

\subsection{Authentication Latency}
\label{sec:latency}

A dedicated campaign of 30 isolated attempts each of ordinary TEAP,
$L{=}3,A{=}1$ full recovery, and $L{=}3,A{=}3$ full recovery, 90 total
and all accepted, measures mean successful-path
EAP-start-to-EAP-success latency at 231.698~ms, 257.096~ms, and
256.558~ms, respectively: the proposed re-verification adds
approximately 25~ms of successful-path EAP latency over ordinary TEAP
in this
\texttt{mac80211\_hwsim} testbed for both full-recovery configurations.

Attempts missing a required active recovery object instead follow the
separate reject path, waiting for the 2000-ms recovery deadline before
rejecting at approximately 2.4--2.5~s, a timing reported qualitatively.

\subsection{Protocol Overhead}
\label{sec:overhead}

Each transport leg's overhead is derived from the implemented message
formats and expressed as $B(L)=B_0+B_1L$ bytes at $L=3$: the TEAP
Batch Request/Response TLVs are $19+18L$/$19+34L$ (73~B/121~B), the $L$
RADIUS recovery VSAs total $50L$ (150~B), and the $L$ Category-127
Action-frame bodies total $49L$ (147~B). Because these quantities occur
on three distinct transport hops, we report them separately.

\section{Discussion and Conclusion}
\label{sec:discussion}

We verify the batched construction, guaranteed $A\ge1$, randomized
positions, litter, and all-round gate across the recovery-model,
trade-off, naive-attacker, and latency campaigns. We derive the
Perception, random-guess, and repeated-observation results under their
independence assumptions.

The current analysis is channel-independent: $r_B$ and $r_E$ are
modeled per-round recovery probabilities, and Eve's retry decision does
not condition on instantaneous channel state. Channel-state-aware
attempt selection, correlated fading, and more general joint
recovery/observation models remain open for physical-layer evaluation.
Physical-layer validation of the modeled recovery asymmetry and
missed-active/litter reception remains open, together with mapping
$L$, $A$, and the recovery codeword to Wi-Fi
modulation-and-coding or finite-blocklength parameters, cryptographic
origin authentication of Category-127 recovery frames,
relay/collusion resistance, real-hardware evaluation, general adaptive
attackers beyond the channel-independent repeated-observation model,
and the cross-session retry policy implementation. The testbed's open
association without RSN or Protected Management Frames is an
evaluation configuration, not a protocol requirement. Outer TEAP
session resumption remains disabled. At $d_p=128$, challenge recovery dominates the mask-guess
contribution to an informed random-guess attacker's success
probability.

The proposed batched PLD re verification
realizes an authentication-oriented step within
EAP-TEAP/RADIUS/IEEE~802.11, implemented end-to-end
in hostap~2.12 and evaluated through complementary full-stack
campaigns alongside analytical informed-attacker,
repeated-observation, and reliability models.

\section*{Acknowledgment}
This work was supported in part by the collaborative NSF-DFG SaTC 2.0 RES project TrustWISC under Grant HA 8970/5-1, and in part by the DFG project TapE under Grant HA 8970/6-1. B. Han (bin.han@rptu.de) is the corresponding author. OpenAI's GPT assisted with debugging of the simulation codes and language polishing of the manuscript.

\bibliographystyle{IEEEtran}
\bibliography{references}

\end{document}

%% file: fig_pld_mechanism.tex
\begin{tikzpicture}[
  >=Latex,
  hdr/.style={font=\footnotesize\bfseries, align=center, text width=3.4cm},
  box/.style={draw, thin, rectangle, align=center, font=\footnotesize,
    fill=white, text width=3.4cm, inner sep=3pt, minimum height=0.7cm},
  note/.style={draw, thin, rectangle, align=center, font=\footnotesize,
    fill=white, text width=10.6cm, inner sep=4pt},
  arr/.style={-{Latex}, thick}
]

\def\xa{0}
\def\xb{4.2}
\def\xc{8.4}

\node[hdr] (ha) at (\xa,0.9) {(a) Active, recovered};
\node[hdr] (hb) at (\xb,0.9) {(b) Active, missed};
\node[hdr] (hc) at (\xc,0.9) {(c) Inactive};

\node[box] (a1) at (\xa,0)    {$p_i,k_i$ fresh, $k_i\ne0$};
\node[box] (b1) at (\xb,0)    {$p_i,k_i$ fresh, $k_i\ne0$};
\node[box] (c1) at (\xc,0)    {$p_i$ only, no $k_i$ used};

\node[box] (a2) at (\xa,-1.3) {$m_i = p_i \oplus k_i$};
\node[box] (b2) at (\xb,-1.3) {$m_i = p_i \oplus k_i$};
\node[box] (c2) at (\xc,-1.3) {$m_i = p_i$};

\node[box] (a3) at (\xa,-2.6) {recovery: \textsc{Valid\_Key}$(\hat{k}_i)$};
\node[box] (b3) at (\xb,-2.6) {recovery: \textsc{No\_Valid\_Key}};
\node[box] (c3) at (\xc,-2.6) {litter $\to$ \textsc{No\_Valid\_Key}};

\node[box] (a4) at (\xa,-3.9) {$\hat{p}_i=m_i\oplus\hat{k}_i=p_i$\\(correct)};
\node[box] (b4) at (\xb,-3.9) {$\hat{p}_i=m_i\ne p_i$\\(deceptive)};
\node[box] (c4) at (\xc,-3.9) {$\hat{p}_i=m_i=p_i$\\(correct)};

\draw[arr] (a1) -- (a2); \draw[arr] (a2) -- (a3); \draw[arr] (a3) -- (a4);
\draw[arr] (b1) -- (b2); \draw[arr] (b2) -- (b3); \draw[arr] (b3) -- (b4);
\draw[arr] (c1) -- (c2); \draw[arr] (c2) -- (c3); \draw[arr] (c3) -- (c4);

\draw[dashed, thin] (b3.south) -- ++(0,-0.35) -| (5.5,-4.55) -| (c3.south);
\node[font=\footnotesize\itshape] at (5.5,-4.75) {same decoder-level outcome: \textsc{No\_Valid\_Key}};

\node[note] (batch) at (4.2,-6.1) {Per attempt: a batch of $L$ rounds,
  exactly $A$ active, $1\le A\le L$ ($A\ge1$ in normal operation); active
  positions chosen uniformly without replacement, no dedicated
  activation flag transmitted; the server accepts only if all $L$
  rounds verify.};

\end{tikzpicture}

%% file: fig_w1v2_flow.tex
\begin{tikzpicture}[
  >=Latex,
  actor/.style={draw, thick, rectangle, minimum width=1.8cm, minimum height=0.8cm, align=center, font=\footnotesize},
  note/.style={draw, thin, rectangle, align=center, font=\footnotesize, fill=white, text width=3.6cm, inner sep=3pt},
  msg/.style={-{Latex}, thick},
  lbl/.style={font=\footnotesize, align=center, text width=2.9cm}
]

\def\sx{0}
\def\ax{4.0}
\def\dx{8.0}

\node[actor] (server) at (\sx,0) {Auth.\\Server};
\node[actor] (ap)     at (\ax,0) {Access\\Point};
\node[actor] (dev)    at (\dx,0) {Device};

\draw[dashed, thin] (server.south) -- ++(0,-11.1);
\draw[dashed, thin] (ap.south)     -- ++(0,-11.1);
\draw[dashed, thin] (dev.south)    -- ++(0,-11.1);

\draw[msg] (\sx,-2.0) -- node[lbl, above, text width=4.3cm] {RADIUS Access-Challenge:\\EAP-Msg\{TEAP Batch Request$[m_1,\dots,m_L]$\}\\+ $L$ VSAs$\{q_1,\dots,q_L\}$} (\ax,-2.0);

\draw[msg] (\ax,-3.9) -- node[lbl, above, text width=3.6cm] {Primary path: 802.11\\Data/EAPOL\{EAP-TEAP/TLS[Batch Request $m_1,\dots,m_L$]\}} (\dx,-3.9);

\draw[msg] (\ax,-5.5) -- node[lbl, above, text width=3.8cm] {Recovery path: $L$ separate\\Category-127 Action frames\\(one per round): $q_1,\dots,q_L$} (\dx,-5.5);

\node[note] (dc) at (\dx,-6.95) {correlate TxnID/Round per round\\decode each $q_i$: \textsc{Valid\_Key}/\textsc{No\_Valid\_Key}\\compute each $\hat{p}_i$; compute each $R_i$};

\draw[msg] (\dx,-8.8) -- node[lbl, above, text width=3.6cm] {Batch Response $\{R_1,\dots,R_L\}$ (TEAP)} (\ax,-8.8);
\draw[msg] (\ax,-9.7) -- node[lbl, above, text width=3.6cm] {Batch Response $\{R_1,\dots,R_L\}$ (RADIUS)} (\sx,-9.7);

\node[note] (sv) at (\sx,-10.5) {verify every Round $1..L$\\occurs exactly once;\\verify every $R_i$};
\draw[msg] (\sx,-11.11) -- node[lbl, above, text width=3.9cm] {EAP success (gated),\\via AP / reject} (\dx,-11.1);

\end{tikzpicture}

%% file: w1v2_campaign_table.tex
\begin{table}[t]
\footnotesize
\caption{Batched recovery-model campaign, 273 attempts.}
\label{tab:w1v2campaign}
\centering
\setlength{\tabcolsep}{3pt}
\begin{tabular}{lccccc}
\hline
\textbf{Mode} & $L$ & $A$ & \textbf{Recov.} & \textbf{Pos.} & \textbf{Accept} \\
\hline
Ordinary TEAP     & -- & -- & --            & -- & 30/30 \\
Full recovery     & 3  & 1  & full          & R  & 30/30 \\
Full recovery     & 3  & 3  & full          & -- & 30/30 \\
Degraded recovery & 3  & 1  & $r=.95$       & F  & 29/30 \\
Degraded recovery & 3  & 1  & $r=.95$       & R  & 29/30 \\
Degraded recovery & 3  & 3  & $r=.95$       & -- & 26/30 \\
Degraded recovery & 3  & 1  & $r=.50$       & F  & 15/30 \\
Degraded recovery & 3  & 1  & $r=.50$       & R  & 15/30 \\
Degraded recovery & 3  & 3  & $r=.50$       & -- & 1/30  \\
\hline
Diagnostic & 3  & 0   & all $q_i$ withheld & -- & 3/3 \\
\hline
\end{tabular}
\vspace{2pt}
\\ {\footnotesize $r$: software recovery-model parameter; F = fixed mask; R = randomized position; $A=0$ diagnostic only.}


\end{table}

%% file: references.bib
@misc{aboba2004eap,
  author       = {Aboba, B. and Blunk, L. and Vollbrecht, J. and Carlson, J. and Levkowetz, H.},
  title        = {Extensible Authentication Protocol ({EAP})},
  howpublished = {IETF RFC 3748},
  year         = {2004},
  month        = jun,
  doi          = {10.17487/RFC3748}
}

@article{poor2017wireless,
  author  = {Poor, H. Vincent and Schaefer, Rafael F.},
  title   = {Wireless Physical Layer Security},
  journal = {Proceedings of the National Academy of Sciences},
  volume  = {114},
  number  = {1},
  pages   = {19--26},
  year    = {2017},
  doi     = {10.1073/pnas.1618130114}
}

@article{xiao2008physical,
  author  = {Xiao, Liang and Greenstein, Larry J. and Mandayam, Narayan B. and Trappe, Wade},
  title   = {Using the Physical Layer for Wireless Authentication in Time-Variant Channels},
  journal = {IEEE Transactions on Wireless Communications},
  volume  = {7},
  number  = {7},
  pages   = {2571--2579},
  year    = {2008},
  doi     = {10.1109/TWC.2008.070194}
}

@article{wang2016physical,
  author  = {Wang, Xianbin and Hao, Peng and Hanzo, Lajos},
  title   = {Physical-Layer Authentication for Wireless Security Enhancement: Current Challenges and Future Developments},
  journal = {IEEE Communications Magazine},
  volume  = {54},
  number  = {6},
  pages   = {152--158},
  year    = {2016},
  month   = jun,
  doi     = {10.1109/MCOM.2016.7498103}
}

@article{xie2021survey,
  author  = {Xie, Ning and Li, Zhuoyuan and Tan, Haijun},
  title   = {A Survey of Physical-Layer Authentication in Wireless Communications},
  journal = {IEEE Communications Surveys \& Tutorials},
  volume  = {23},
  number  = {1},
  pages   = {282--310},
  year    = {2021},
  month   = {First Quarter},
  doi     = {10.1109/COMST.2020.3042188}
}

@misc{dekok2026teap,
  author       = {DeKok, A.},
  title        = {Tunnel Extensible Authentication Protocol ({TEAP}) Version 1},
  howpublished = {IETF RFC 9930},
  year         = {2026},
  month        = feb
}

@misc{rigney2000radius,
  author       = {Rigney, C. and Willens, S. and Rubens, A. and Simpson, W.},
  title        = {Remote Authentication Dial In User Service ({RADIUS})},
  howpublished = {IETF RFC 2865},
  year         = {2000},
  month        = jun
}

@misc{aboba2003radiuseap,
  author       = {Aboba, B. and Calhoun, P.},
  title        = {{RADIUS} (Remote Authentication Dial In User Service) Support For Extensible Authentication Protocol ({EAP})},
  howpublished = {IETF RFC 3579},
  year         = {2003},
  month        = sep
}

@standard{ieee8021x2020,
  title        = {{IEEE} Standard for Local and Metropolitan Area Networks -- Port-Based Network Access Control},
  organization = {IEEE},
  howpublished = {IEEE Std 802.1X-2020},
  year         = {2020},
  month        = feb
}

@standard{ieee80211_2024,
  title        = {{IEEE} Standard for Information Technology -- Telecommunications and Information Exchange between Systems -- Local and Metropolitan Area Networks -- Specific Requirements -- Part 11: Wireless {LAN} Medium Access Control ({MAC}) and Physical Layer ({PHY}) Specifications},
  organization = {IEEE},
  howpublished = {IEEE Std 802.11-2024},
  year         = {2025},
  month        = apr
}

@article{shan2013phycram,
  author  = {Shan, Dan and Zeng, Kai and Xiang, Weidong and Richardson, Paul and Dong, Yan},
  title   = {{PHY-CRAM}: Physical Layer Challenge-Response Authentication Mechanism for Wireless Networks},
  journal = {IEEE Journal on Selected Areas in Communications},
  volume  = {31},
  number  = {9},
  pages   = {1817--1827},
  year    = {2013},
  month   = sep,
  doi     = {10.1109/JSAC.2013.130914}
}

@inproceedings{han2025semantic,
  author    = {Han, Bin and Zhu, Yao and Schmeink, Anke and Caire, Giuseppe and Schotten, Hans D.},
  title     = {A Semantic Model for Physical Layer Deception},
  booktitle = {2025 IEEE International Conference on Communications (ICC)},
  address   = {Montreal, QC, Canada},
  pages     = {4129--4134},
  year      = {2025},
  doi       = {10.1109/ICC52391.2025.11162021}
}

@inproceedings{han2023nom,
  author    = {Han, Bin and Zhu, Yao and Schmeink, Anke and Schotten, Hans D.},
  title     = {Non-Orthogonal Multiplexing in the {FBL} Regime Enhances Physical Layer Security with Deception},
  booktitle = {2023 IEEE 24th International Workshop on Signal Processing Advances in Wireless Communications (SPAWC)},
  pages     = {211--215},
  year      = {2023},
  doi       = {10.1109/SPAWC53906.2023.10304549}
}

@article{chen2025twc,
  author  = {Chen, Wanli and Han, Bin and Zhu, Yao and Schmeink, Anke and Caire, Giuseppe and Schotten, Hans D.},
  title   = {Physical Layer Deception With Non-Orthogonal Multiplexing},
  journal = {IEEE Transactions on Wireless Communications},
  volume  = {24},
  number  = {5},
  pages   = {3923--3936},
  year    = {2025},
  month   = may,
  doi     = {10.1109/TWC.2025.3535545}
}

@inproceedings{chen2025ofdm,
  author    = {Chen, Wanli and Han, Bin and Zhu, Yao and Schmeink, Anke and Schotten, Hans D.},
  title     = {Physical Layer Deception in {OFDM} Systems},
  booktitle = {2025 Joint European Conference on Networks and Communications \& 6G Summit (EuCNC/6G Summit), Security Aspects and Quantum Communications Track},
  pages     = {715--720},
  year      = {2025},
  doi       = {10.1109/EUCNC/6GSUMMIT63408.2025.11036978}
}
